\documentclass[a4paper,12pt]{article}
\pdfoutput = 1
\usepackage{jheppub}
\usepackage{amsmath,amssymb,amscd,braket,amsfonts}
\usepackage{color}
\usepackage{graphicx}
\usepackage{slashed}
\usepackage{amsthm}

\usepackage{lettrine, Romantik}

\newcommand{\bea}{\begin{eqnarray}}
\newcommand{\eea}{\end{eqnarray}}
\newcommand{\be}{\begin{equation}}
\newcommand{\ee}{\end{equation}}
\newcommand{\ba}{\begin{align}}
\newcommand{\ea}{\end{align}}

\newcommand{\im}{\mathfrak{Im}}
\newcommand{\re}{\mathfrak{Re}}

\title{Analytic structure of holographic thermal correlators from Fourier series} 

\author{Paolo Arnaudo and}
\author{Benjamin Withers}
\affiliation{Mathematical Sciences and STAG Research Centre, University of Southampton, Highfield, Southampton SO17 1BJ, UK}
\emailAdd{p.arnaudo@soton.ac.uk}
\emailAdd{b.s.withers@soton.ac.uk}

\abstract{
We compute the holographic Euclidean two-point function of scalar operators in a thermal state. We work directly using the Fourier series on the thermal circle. The Fourier series does not converge as a function, but instead converges as a distribution, consistent with QFT expectations. The result is manifestly periodic and consistent with analyticity in the strip $0<\mathfrak{Re}(\tau)<\beta$. Expanding in $\tau$ we obtain all OPE coefficients, including the double-trace sector. Thus our approach has an advantage compared to recent work where double-traces were bootstrapped from stress-tensor data. Bouncing singularities appear as non-perturbative sectors in the transseries for Fourier coefficients, but their transseries parameters are all zero in the case of the Euclidean correlator.
}

\begin{document}
\maketitle

\section{Introduction}
Holography \cite{Maldacena:1997re, Witten:1998qj} provides a practical computational framework in which strongly coupled quantum field theories at finite temperature are dual to asymptotically AdS black hole spacetimes \cite{Witten:1998zw}. In this context, correlation functions of local operators are computable through black hole perturbation theory, making thermal physics at strong coupling directly accessible.

In this work we consider the Euclidean correlator of identical scalar operators at finite temperature $T = \beta^{-1}$ on $S^1_\beta\times \mathbb{R}^{d-1}$,
\be
G_E(\tau,x) = \left<\mathcal{O}(\tau,x)\mathcal{O}(0,0)\right>_\beta, \label{GE}
\ee
where the scalars have generic conformal dimensions $\Delta >d/2$. In the holographic context it is interesting to consider any signatures of the black hole spacetime that might be present in this observable, particularly evidence of the black hole singularity when analytically continuing to complex $\tau$ \cite{Fidkowski:2003nf, Festuccia:2005pi}. 

Much recent interest has centred on the operator product expansion (OPE) of \eqref{GE}. For large $N$ theories there are only two sectors in the OPE coming from multi stress-tensors and double-trace operators,
\be
G_E(\tau,x) = g_{T}(\tau, x) + g_{[\phi\phi]}(\tau, x).
\ee
The stress tensor sector $g_{T}(\tau, x)$ (in which we include the identity) is easily accessible from holographic computations, where its expansion coefficients can be obtained from an asymptotic expansion in large momenta \cite{Fitzpatrick:2019zqz}.
In particular, in earlier work by one of us \cite{Parisini:2023nbd}, it was shown that the small $\tau$ expansion of $g_{T}(\tau, 0)$ has a finite radius of convergence in the complex $\tau$ plane, set by singularities at
\be
\tau_j = \frac{\beta}{\sqrt{2}}e^{\frac{i\pi}{4}(1 + 2j)}, \qquad j = 1,2,3,4, \label{bouncingtau}
\ee
corresponding to spacelike geodesics which bounce off the black hole singularity \cite{Fidkowski:2003nf, Festuccia:2005pi}. This result was obtained at $d=4, \Delta=3/2$ in \cite{Parisini:2023nbd} and was explored further in \cite{Ceplak:2024bja} for general $\Delta$, and extended to charged black holes with timelike singularities in \cite{Ceplak:2025dds}. In \cite{Buric:2025anb, Buric:2025fye} it was shown that imposing periodicity through thermal image sums, or using a KMS bootstrap approach, the double traces $g_{[\phi\phi]}(\tau, x)$ are correctly recovered from the stress tensor data. The expectation is that the singularities of $g_{T}(\tau, x)$ (of which \eqref{bouncingtau} are the closest to the origin) are cancelled by those of $g_{[\phi\phi]}(\tau, x)$, restoring analyticity in the strip $0 < \re(\tau) < \beta$.\footnote{This step does not occur for the retarded Green's function, where the double-trace sector is absent \cite{Afkhami-Jeddi:2025wra, Giombi:2026kdz}.} This connection has also been explored from a field theory thermal bootstrap perspective in \cite{Iliesiu:2018fao, Barrat:2025nvu, Barrat:2025twb, Niarchos:2025cdg}.

In this work we adopt a different strategy, whereby we compute the correlator directly, rather than bootstrapping it from stress-tensor OPE data. Previous approaches to this problem involved numerically solving PDEs \cite{Parisini:2023nbd, Buric:2025fye, Barrat:2025twb}. Here we focus on the correlator evaluated at fixed spatial momentum $k$,
\be
G_E(\tau, k) = \int d^{d-1} x\, \left<\mathcal{O}(\tau,x)O(0,0)\right>_\beta\, e^{-i k x}. \label{spatialFT}
\ee
In particular, we will work with its representation as a Fourier series on the thermal circle,
\be
G_E(\tau, k) = \frac{1}{\beta}\sum_{n=-\infty}^\infty   \widetilde{G}_E(\zeta_n, k)e^{i \zeta_n \tau},\qquad \zeta_n = \frac{2\pi n}{\beta}. \label{fourierseries}
\ee
This is a natural representation that emerges from the bulk PDE problem after separation of variables, and the coefficients are straightforwardly computed by solving radial ODEs. Our goal will be to use this representation to compute the correlator for $\tau \in \mathbb{R}$, its analytic continuation to $\tau \in \mathbb{C}$, and to read off OPE coefficients from the result. In what follows we restrict our attention to $d=4$.

The Fourier sum \eqref{fourierseries} does not converge in the sense of functions. Instead, as can be seen from the power-law growth of the coefficients,\footnote{Here, and throughout this paper `$\sim$' means `asympotic to', i.e. including the correct coefficients.}
\be
\widetilde{G}_E(\zeta_n, k) \sim -\frac{2\pi \csc(\pi\Delta)}{\Gamma(\Delta-2)^2} \left(\frac{\pi}{\beta}\right)^{2\Delta -4} n^{2\Delta-4}, \qquad n\to \infty \label{introasygrowth}
\ee
the sum converges in the sense of distributions. In particular, the $n>0$ sum converges for $\im\,\tau>0$ to an analytic function, similarly, the $n<0$ sum converges for $\im\, \tau<0$ to an analytic function, and each converges on their common boundary $\im\,\tau=0$ as a distribution. Meaning that whilst \eqref{fourierseries} does not converge pointwise for $\tau \in \mathbb{R}$, instead, the sequence of partial sums of the Fourier series integrated against test functions, $\varphi \in C^\infty(S^1)$, does,
\be
\lim_{N\to\infty}\left<\frac{1}{\beta}\sum_{n=-N}^N   \widetilde{G}_E(\zeta_n, k)e^{i \zeta_n \tau}, \varphi\right>  = \left<G_E(\tau, k), \varphi\right>,\label{fouriersumtestfunction}
\ee
where the angle brackets here mean $\left<f, \varphi\right> = \int_0^{\beta} f(\tau)\varphi(\tau)d\tau$.\footnote{Here, adding finite local counterterms in the sense of holographic renormalisation \cite{Skenderis:2002wp} contributes analytic pieces to $\widetilde{G}_E(\zeta_n, k)$, and thus contact terms in the form of deltas and derivatives of deltas to the distribution. Thus evaluating \eqref{fouriersumtestfunction} using a test function with support at the singular point will be correspondingly scheme dependent. Nevertheless, given a choice of holographic counterterms, such terms are determined.} In practical terms, adopting a suitable regularisation method will allow us to provide an accurate computation of finite local data for $G_E(\tau,k)$ away from singularities, including in the complex $\tau$ plane after analytic continuation. See also \cite{Kravchuk:2020scc, Kravchuk:2021kwe} for a discussion of the distributional nature of correlators in the context of the CFT bootstrap axioms.

The representation \eqref{fourierseries} is manifestly periodic in $\tau \sim \tau +\beta$. Expanding the result at small $\tau$ automatically contains the double trace OPE sector. The small $\tau$ expansion at fixed $k$ differs from the OPE expansion at fixed $x$, due to the spatial integral \eqref{spatialFT}. For $k=0$ and generic $\Delta$ it takes the form
\bea
G_E(\tau, k=0) &=& \frac{1}{\tau^{2\Delta -3}}\sum_{m=0}^\infty a_m \left(\frac{\tau}{\beta}\right)^{4m} + \frac{1}{\beta^{2\Delta-3}}\sum_{q=0}^\infty b_q \left(\frac{\tau}{\beta}\right)^{2q},\label{OPEintro}
\eea
while for integer values of $2\Delta$ logarithms can appear. The $a_m$ correspond to the stress-tensor sector and are determined analytically by the asymptotic $1/n$ expansion of the Fourier coefficients. The $b_q$ are also directly determined from the same set of coefficients. The final expressions are given by \eqref{acoeffs} and \eqref{bcoeffs} respectively. In particular, no thermal bootstrap or image sums are required to compute them. 

Finally let us comment on the three techniques we use to evaluate the Fourier coefficients $\widetilde{G}_E(\zeta_n, k)$ in holography. In the case of $d=4$ scalar perturbations they are governed by the Heun differential equation, and evaluating the ratios of connection coefficients between local solutions of the Heun equation yields $\widetilde{G}_E(\zeta_n, k)$. In the first instance we utilise numerics to obtain the connection coefficients through evaluation of Heun functions and their derivatives. Second we utilise an analytic approximation method in which the connection coefficients are given as series expansions in a parameter $X$, known as the instanton expansion from the Seiberg-Witten geometry viewpoint \cite{seiberg1994, seiberg1994a, Aminov:2020yma, Bonelli:2022ten, Lisovyy2022}. The third method involves obtaining $\widetilde{G}_E(\zeta_n, k)$ as an asymptotic expansion in $1/n$, which can be computed analytically using the approach of \cite{Fitzpatrick:2019zqz}, and which then requires a resurgence analysis to complete. In other words, in our work, the technique \cite{Fitzpatrick:2019zqz} which is usually used to compute stress-tensor OPE coefficients, here instead applies to the asymptotic Fourier series coefficients, and thus to the manifestly periodic result. We utilise all three approaches and find consistent final picture between them for $G_E(\tau, k)$.
\\\\
The paper is organised as follows. In section \ref{sec:laplace} we discuss a warm-up example of the Laplace equation on a disk with a delta function boundary condition, to highlight the distributional nature of the response. In section \ref{sec:fouriercoefficients} we detail the computation of Fourier series coefficients in holography. In section \ref{sec:analyticity} we show how the Fourier series data can be used to construct the complex $\tau$-plane after suitable regularisation, where we find results consistent with analyticity in the strip $0 < \re(\tau) < \beta$. In section \ref{sec:OPE} we extract double-trace OPE coefficients from asymptotics of the Fourier series, and demonstrate precise numerical agreement with the other approaches. We finish with an outlook in section \ref{sec:outlook}.
\\\\
\emph{Attached to the arXiv submission, we include the {\tt Mathematica} file ``ancillary.nb'' with the first 70 coefficients of the asymptotic $1/n$ expansion of $\widetilde{G}_E(\zeta_n,k=0)$.}

\section{A warm up example: Laplace in a disk}\label{sec:laplace}

As an introduction to the distributional nature of the thermal correlators, in this section we analyse a toy problem, provided by the Laplace differential operator.
Let us consider on the unit disk, $dr^2 + r^2 d\theta^2$, the differential problem with Dirichlet boundary condition
\be
\nabla^2\phi = 0, \qquad \phi\big|_{r=1} = \delta(\theta).\label{laplaceBVP}
\ee
As an analogue to a thermal correlator, we consider the boundary normal derivative of the resulting solution,
\be
G_E \equiv \lim_{r\to 1^-} \partial_r \phi.
\ee
Here $\theta$ plays the role of the Euclidean time coordinate, with period $2\pi$. For $r<1$, the solution to \eqref{laplaceBVP} is given by the \emph{Poisson kernel}
\be\label{poissonkernel}
\phi = \frac{1-r^2}{2\pi(1-2r\cos\theta + r^2)},
\ee
whose decomposition in Fourier modes is given by
\begin{equation}\label{fourierpoissonkernel}
\phi=\frac{1}{2\pi}\sum_{k=-\infty}^{\infty}r^{|k|}e^{i\,k\,\theta},
\end{equation}
from which we get the decomposition of $G_E$ in Fourier modes, 
\begin{equation}\label{kernelDtNfourier}
G_E=\frac{1}{2\pi}\sum_{k=-\infty}^{\infty}|k|e^{i\,k\,\theta}.
\end{equation}
The Fourier series \eqref{fourierpoissonkernel} converges to the function \eqref{poissonkernel} for $r<1$, while \eqref{kernelDtNfourier} converges in the sense of distributions. Aside from the Fourier series itself, there are different ways to represent this distribution.

The problem is well-studied in the mathematics literature, where $G_E$ is known as the \emph{kernel of the Dirichlet-to-Neumann (DtN) operator} \cite{taylor2018dirichlettoneumann, girouard2022dirichlet}. 
The DtN operator on the unit disk is the linear map
\begin{equation}
\Lambda : C^\infty(S^1) \to C^\infty(S^1), 
\end{equation}
defined by its action on the Fourier basis:
\begin{equation}
\Lambda \frac{e^{i k \theta}}{2\pi} = \frac{1}{2\pi}|k| \, e^{i k \theta}, \quad k \in \mathbb{Z}.
\end{equation}
This is an elliptic pseudodifferential operator, and its kernel is defined as a distribution on $S^1\times S^1$
\begin{equation}
G_E \in \mathcal{D}'(S^1 \times S^1), 
\end{equation}
defined by the pairing
\begin{equation}
(\Lambda f)(\theta) = \int_{S^1}G_E(\theta,\theta')\,f(\theta')\mathrm{d}\theta',
\end{equation}
for every $f\in C^\infty(S^1)$.
Considering the decomposition in Fourier modes, this action is realised by 
\begin{equation}
G_E(\theta,\theta') = \frac{1}{2\pi}\sum_{k \in \mathbb{Z}} |k| \, e^{i k (\theta - \theta')} \in \mathcal{D}'(S^1 \times S^1).
\end{equation}
Since the kernel depends only on the difference $\theta-\theta'$, one can consider instead the single-variable distribution $G_E(\theta-\theta')$, as we do in this work, recovering \eqref{kernelDtNfourier}.

A particularly illustrative representation is in terms of principal value distributions,
\be
G_E = -\frac{1}{4\pi}\,\text{p.v.}\left(\frac{1}{\sin^2\!\left(\frac{\theta}{2}\right)}\right),
\ee
where $\text{p.v.}$ is understood in the distributional sense as a principal value of order two, more precisely, defined by
\begin{equation}
\text{p.v.}\,\csc^2\,\left(\frac{\theta}{2}\right)\equiv-2\,\partial_\theta\,\left[\text{p.v.}\,\cot\,\left(\frac{\theta}{2}\right)\right].
\end{equation}
In particular, this is equivalent to the integral representation against test functions $\varphi \in C^\infty(S^1)$,
\be
\left<G_E,\varphi\right> = \frac{1}{2\pi}\,\text{p.v.}\,\int_{0}^{2\pi}\cot\!\left(\frac{\theta}{2}\right)\,\partial_{\theta}\varphi(\theta)\,d\theta,\label{laplaceGEvarphi}
\ee
where now $\text{p.v.}$ denotes the Cauchy principal value integral. From this representation we see that $G_E$ is represented by a function away from the singular points,
\be
-\frac{1}{4\pi}\frac{1}{\sin^2\!\left(\frac{\theta}{2}\right)}.\label{GEfunc}
\ee
However understanding $G_E$ as a distribution and not as an ordinary function is essential for obtaining the correct integrals against test functions, in any case where the test function has support at $\theta = 0$. Moreover we emphasise that even though $G_E$ is described by \eqref{GEfunc} away from the singular points, the Fourier series does not converge there. Thus, it is instructive to consider further representations of the distribution which will allow us to access \eqref{GEfunc} from the Fourier series directly.

One convenient way to make use of the distributional Fourier series \eqref{kernelDtNfourier} is to introduce an $\mathrm{i}\epsilon$ prescription. Concretely, for $\epsilon>0$ we define the regulated kernel by damping the high Fourier modes,
\begin{equation}\label{eq:GE-eps-def}
G_\epsilon(\theta)\equiv\frac{1}{2\pi}\sum_{k\in\mathbb Z} |k|\,e^{-|k|\epsilon}\,e^{ik\theta}.
\end{equation}
For every fixed $\epsilon>0$ the series is absolutely and uniformly convergent, hence $G_\epsilon\in C^\infty(S^1)$. In particular, away from the singular points one recovers \eqref{GEfunc} in the limit $\epsilon\to 0^+$.
More generally, $G_\epsilon$ converges to $G_E$ in the sense of distributions as $\epsilon\to 0^+$: for any test function $\varphi\in C^\infty(S^1)$, writing 
\begin{equation}
\varphi(\theta)=\sum_{k\in\mathbb Z}\hat{\varphi}_k\, e^{ik\theta},
\end{equation}
we have
\begin{equation}
\langle G_\epsilon, \varphi\rangle=\sum_{k\in\mathbb Z}|k|\,e^{-|k|\epsilon}\,\hat \varphi_{-k}\xrightarrow{\ \epsilon\to 0^+\ }\sum_{k\in\mathbb Z}|k|\,\hat \varphi_{-k}=\langle G_E, \varphi\rangle.
\end{equation}
Thus, the $\mathrm{i}\epsilon$ prescription realises $G_E$ as an Abel-summation limit of its Fourier series.
In this case, this regularisation is equivalent to moving slightly inside the disk, and using the Abel regulator already appearing in \eqref{fourierpoissonkernel}. Indeed, by choosing $r=e^{-\epsilon}$, one has $r^{|k|}=e^{-|k|\epsilon}$, which exponentially suppresses high frequencies.

Another convenient representation uses Pad\'e approximants to regularise the divergent Fourier series. This is one of the approaches we adopt later in the holographic computation because it allows us to analytically continue further into the complex $\tau$ plane. Writing $z=e^{i\theta}$, the Fourier series becomes a Laurent series
\begin{equation}
G_E(\theta)= \frac{1}{2\pi}\sum_{k\in\mathbb Z}|k|\,z^k,
\end{equation}
which naturally splits into two power series:
\begin{equation}
\begin{aligned}
G_E^{(+)}(z)&=\frac{1}{2\pi}\sum_{k\ge 1}k\,z^k \quad |z|<1,\\ 
G_E^{(-)}(z)&=\frac{1}{2\pi}\sum_{k\ge 1}k\,z^{-k} \quad |z|>1.
\end{aligned}
\end{equation}
For every $N\in\mathbb{N}$, one can then form diagonal Padé approximants of $G_E^{(\pm)}(z)$, that is, rational functions $R_N^{(\pm)}(z)$ whose Taylor expansions match the corresponding series to order $2N$.
The regulated kernel is obtained by evaluating these rational functions on the unit circle as boundary values,
\begin{equation}
G_{N}(\theta)\equiv R_N^{(+)}(e^{i\theta})+R_N^{(-)}(e^{i\theta}).
\end{equation}
In this way, Padé provides a rational analytic continuation of the Fourier data, and taking the distributional limit $N\to\infty$ recovers the same distribution $G_E\in\mathcal D'(S^1)$.

\section{Fourier coefficients from holography} \label{sec:fouriercoefficients}
In this section we outline the computation of the Fourier coefficients in the series \eqref{fourierseries} in holography, for the case $d=4$ dual to the Euclidean planar Schwarzschild-AdS$_5$ metric,
\bea
ds^2 &=& f(r) d\tau^2 + \frac{dr^2}{f(r)} + r^2 d\vec{x}_3^2,\\
f(r) &=& r^2 - \frac{r_h^4}{r^2},
\eea
where $r_h$ is the location of the horizon, and the temperature is given by
\begin{equation}
T=\beta^{-1}=\frac{r_h}{\pi}.
\end{equation}
The scalar operator $\mathcal{O}$ with dimension $\Delta$ is dual to a scalar field with equation of motion $(\nabla^2 -m^2)\Phi = 0$ where $m^2 = \Delta(\Delta - 4)$. Utilising separation of variables,
\be
\Phi(\tau, r, \vec{x}) = e^{i \vec{k}\cdot \vec{x}}\Phi_k(\tau, r) = e^{-\omega \tau + i \vec{k}\cdot \vec{x}} \phi(r)
\ee
gives the following radial ODE,
\begin{equation}\label{radialplanar}
\frac{1}{r^3}\frac{\mathrm{d} }{\mathrm{d} r}\left(r^3 f(r) \frac{\mathrm{d} \phi(r)}{\mathrm{d} r}\right)+\left(\frac{\omega ^2}{f(r)}-\frac{k^2}{r^2}-\Delta(\Delta-4)\right)\phi(r)=0,
\end{equation}
where we used the notation $k=|\vec{k}|$.

Working at fixed $k$, we consider modes at Matsubara frequencies $\omega=i\,\zeta_n$, and for each $n$ we will build a solution that is normalisable at the boundary and regular in the interior, $\phi_{nk}(r)$. As $r\to \infty$ we normalise it so that $\phi_{nk}(r) \sim \beta^{-1} r^{\Delta-4}$. Then, 
\be
\Phi_k(\tau,r) = \sum_{n=-\infty}^\infty e^{i \zeta_n \tau} \phi_{nk}(r) \sim \delta(\tau) r^{\Delta - 4},
\ee
i.e. it gives a unit strength delta function for boundary Dirichlet data at $\tau = 0$. Then, the Euclidean Green's function is read off from the normalisable data, 
\be\label{FourierSum}
G_E(\tau, k) = \sum_{n=-\infty}^\infty e^{i \zeta_n \tau} (2\Delta-4)\phi^{(\Delta)}_{nk}.
\ee
Thus, by comparison with \eqref{fourierseries} we have the expression for Fourier coefficients as the ratio of normalisable and non-normalisable data
\be
\widetilde{G}_E(\zeta_n, k) = (2\Delta - 4)\frac{\phi^{(\Delta)}_{nk}}{\phi^{(4-\Delta)}_{nk}}. \label{asydataratio}
\ee
This can be written in terms of connection coefficients relating the local solutions around the horizon to the ones around the AdS boundary, which we construct in the following subsection. For later use we note the special case $\omega = k = 0$ which can be solved directly, to find
\be
\widetilde{G}_E(0,0) = -\frac{4\Gamma\left(2-\frac{\Delta}{2}\right)\Gamma\left(\frac{\Delta}{4}\right)^2}{\Gamma\left(1-\frac{\Delta}{4}\right)^2\Gamma\left(-1+\frac{\Delta}{2}\right)}\left(\frac{\pi}{\beta}\right)^{2\Delta - 4}. \label{G00}
\ee

\subsection{Evaluating Heun connection coefficients}
The Fourier coefficient \eqref{asydataratio} can be expressed in term of connection coefficients of local solutions to the Heun equation. By defining
\begin{align}
v&=\frac{r^2-r_h^2}{2r^2},\\
\phi_{n\,k}(r)&=p(v)\,\psi_{n\,k}(v),\\
p(v)&=\frac{\sqrt{1-2 v}}{\sqrt{v\,(1-v)}},
\end{align}
the wave function $\psi_{n\,k}(v)$ satisfies a Heun equation
\begin{equation}\label{heunnormalform}
\begin{aligned}
&\Biggl[\frac{\mathrm{d}^2}{\mathrm{d}v^2} + \frac{\frac{1}{4}-a_0^2}{v^2}+\frac{\frac{1}{4}-a_1^2}{(v-1)^2} + \frac{\frac{1}{4}-a_{X}^2}{(v-X)^2}- \frac{\frac{1}{2}-a_1^2 -a_{X}^2 -a_0^2 +a_\infty^2 + u}{v(v-1)}+\frac{u}{v(v-X)} \Biggr]\psi_{n\,k}(v)=0,
\end{aligned}
\end{equation}
with dictionary
\begin{equation}\label{Heundictio}
\begin{aligned}
X &=\frac{1}{2},\qquad a_0= -\frac{i\omega}{4 r_h},\qquad a_1= \frac{\omega}{4r_h},\\
a_X&= \frac{\Delta -2}{2},\qquad a_{\infty}=0,\qquad u= \frac{k^2-\omega ^2}{4 r_h^2},
\end{aligned}
\end{equation}
where again $\omega=i\,\zeta_n$.
In the $v$ variable, the AdS boundary is at $v=X$ and the horizon is at $v=0$.
The mapping of the scalar perturbation of the AdS$_{d+1}$ black brane with $d=4$ was already analysed in \cite{Dodelson:2022yvn} with a different choice of variable and dictionary. 

In terms of the parameters in \eqref{Heun parameters}, the four relevant local solutions are as follows:
\begin{itemize}
\item The non-normalisable solution at large $r$ behaving as $\phi_{n k}^{N.N.}(r)\sim 2^{\frac{\Delta -1}{2}}\left(\frac{r}{r_h}\right)^{\Delta-4}$ is
\begin{equation}
\begin{aligned}
\phi_{n k}^{N.N.}(r)=&\,p(v)\,v^{\gamma /2} (1-v)^{\delta /2}X^{-\gamma /2} (1-X)^{-\delta /2} (X-v)^{\epsilon /2}\\
&\text{Heun}\left(\frac{X}{X-1},\frac{q-\alpha_1  \alpha_2  X}{1-X},\alpha_1 ,\alpha_2 ,\epsilon ,\delta ,\frac{v-X}{1-X}\right),
\end{aligned}
\end{equation}

\item The normalisable solution at large $r$ behaving as $\phi_{n k}^{N.}(r)\sim 2^{\frac{3-\Delta }{2}}\left(\frac{r}{r_h}\right)^{-\Delta}$ is
\begin{equation}
\begin{aligned}
\phi_{n k}^{N.}(r)=\,&p(v)\,v^{\gamma /2} (1-v)^{\delta /2}X^{-\gamma /2} (1-X)^{-\delta /2} (X-v)^{1-\epsilon /2}\\
&\text{Heun}\Bigl(\frac{X}{X-1},\frac{-q+X\,(\alpha_2-\gamma-\delta)\,(\alpha_1-\gamma-\delta)+\gamma\,(\alpha_1+\alpha_2-\gamma-\delta)}{X-1},\\
&\qquad-\alpha_1+\gamma+\delta,-\alpha_2+\gamma+\delta,2-\epsilon,\delta,\frac{v-X}{1-X}\Bigr),
\end{aligned}
\end{equation}

\item The regular solution at $r=r_h$ when $\omega = i\,\zeta_n$ with $n<0$, behaving as $\phi_{n k}^{<}(r)\sim \left[r_h^3\,(r-r_h)\right]^{-n/2}$ is
\begin{equation}
\phi_{n k}^{<}(r)=p(v)v^{\gamma/2}(1-v)^{\delta/2}(X-v)^{\epsilon/2}X^{-\epsilon/2}\text{Heun}(X,q,\alpha_1,\alpha_2,\gamma,\delta,v),
\end{equation}

\item The regular solution at $r=r_h$ when $\omega = i\,\zeta_n$ with $n>0$, behaving as $\phi_{n k}^{>}(r)\sim \left[r_h^3\,(r-r_h)\right]^{n/2}$ is
\begin{equation}
\begin{aligned}
\phi_{n k}^{>}(r)=&\,p(v)v^{1-\gamma/2}(1-v)^{\delta/2}(X-v)^{\epsilon/2}X^{-\epsilon/2}\\
&\text{Heun}(x,q-(\gamma-1)(X\,\delta+\epsilon),\alpha_1+1-\gamma,\alpha_2+1-\gamma,2-\gamma,\delta,v).
\end{aligned}
\end{equation}

\end{itemize}

We can now introduce explicitly the connection formulae relating these local solutions.
Let us start from the case $n>0$. We write
\begin{equation}\label{connectionn>0}
\psi_{n\,k}^{>}(v)=\mathcal{C}_{>,N.}\,\psi_{n\,k}^{N.}(v)+\mathcal{C}_{>,N.N.}\,\psi_{n\,k}^{N.N.}(v),
\end{equation}
where, using the notation
\begin{equation}
W\left[g_1(v),g_2(v)\right]=g_1(v)\,g'_2(v)-g'_1(v)\,g_2(v),
\end{equation}
we have
\begin{equation}\label{connformulas}
\begin{aligned}
\mathcal{C}_{>,N.}&=\frac{W[\psi^>_{n\,k}(v),\psi^{N.N.}_{n\,k}(v)]}{W[\psi^{N.}_{n\,k}(v),\psi^{N.N.}_{n\,k}(v)]}=\frac{W[\psi^>_{n\,k}(v),\psi^{N.N.}_{n\,k}(v)]}{2a_X},\\
\mathcal{C}_{>,N.N.}&=\frac{W[\psi^>_{n\,k}(v),\psi^{N.}_{n\,k}(v)]}{W[\psi^{N.N.}_{n\,k}(v),\psi^{N.}_{n\,k}(v)]}=-\frac{W[\psi^>_{n\,k}(v),\psi^{N.}_{n\,k}(v)]}{2a_X}.
\end{aligned}
\end{equation}
For $n<0$, the same formulae hold by replacing $>$ with $<$.

With these notations, the Fourier coefficients in \eqref{FourierSum} are given by 
\begin{equation}\label{tildeGEplanar}
\widetilde{G}_E(\zeta_n, k) = (2\Delta-4)\frac{\phi_{nk}^{(\Delta)}}{\phi_{nk}^{(4-\Delta)}}=\frac{2\Delta-4}{r_h^{4-2\Delta}}\times \begin{cases}
    \frac{\mathcal{C}_{<,N.}}{\mathcal{C}_{<,N.N.}} & n<0\\
    \frac{\mathcal{C}_{>,N.}}{\mathcal{C}_{>,N.N.}}  & n>0
\end{cases}.
\end{equation}
The connection coefficients can be computed numerically using the Wronskian between the local solutions, but they also admit analytic expressions, which are given as series expansions in the parameter $X$, known as the instanton expansion from the Seiberg-Witten geometry viewpoint \cite{seiberg1994, seiberg1994a, Aminov:2020yma, Bonelli:2022ten, Lisovyy2022}. In this work we will make use of both approaches. The former is useful when accurate coefficients are required, the latter is useful for getting a broad handle on the analytic structure, since in this case the convergence in $X$ is slow.

By using the expressions reported in Appendix \ref{app:gauge}, and with the substitution $\omega=i\zeta_n$, the Green's function for $n>0$ reads
\begin{equation}\label{Fouriercoeffgauge}
\widetilde{G}_E(\zeta_{n>0},k)=\frac{2\Delta-4}{r_h^{4-2\Delta}}X^{2-\Delta }e^{-\partial_{a_X}F(X)} \frac{\Gamma (2-\Delta )  \Gamma \left(\frac{n}{2}-a+\frac{\Delta -1}{2}\right) \Gamma \left(\frac{n}{2}+a+\frac{\Delta -1}{2}\right)}{\Gamma (\Delta -2) \Gamma \left(\frac{n}{2}-a +\frac{3-\Delta}{2}\right) \Gamma \left(\frac{n}{2}+a+\frac{3-\Delta}{2}\right)},
\end{equation}
where the quantities $a$ and $F(X)$ are defined in Appendix \ref{app:gauge}.
Moreover, $\widetilde{G}_E(\zeta_{n<0},k)=\widetilde{G}_E(\zeta_{(-n)>0},k)$.

\subsection{Asymptotic expansion in $1/n$} \label{sec:1onnmethod}

In this subsection, we solve the differential equation \eqref{radialplanar} perturbatively in the large $\omega$ expansion, or, equivalently, in the large $n$ expansion, at $k=0$.\footnote{We remark that fixing $k=0$ helps simplifying the result, but the procedure works similarly for any value of $k$.} We follow the procedure outlined in Appendix A of \cite{Afkhami-Jeddi:2025wra}. First, we define the variable $y=\omega/r$, and we redefine the wave function in \eqref{radialplanar} as
\begin{equation}
\phi(r)=\frac{1}{\sqrt{\left(\frac{r}{r_h}\right)^4-1}}\chi(y).
\end{equation}
The resulting differential equation satisfied by $\chi(y)$ reads
\begin{equation}
\begin{aligned}
&y^2\,\chi''(y)+y\,\chi'(y)+\left(y^2-(\Delta -2)^2\right)\chi(y)\\
&-y^4\,\frac{r_h^4}{\omega ^4} \left(\frac{(\Delta -2)^2-y^2}{1-y^4\,\frac{r_h^4}{\omega ^4}}-\frac{y^2+4}{\left(1-y^4\,\frac{r_h^4}{\omega ^4}\right)^2}\right)\chi(y)=0,
\end{aligned}
\end{equation}
where the second line encodes higher-order terms in the large $\omega$ expansion.

We solve the differential equation perturbatively in the small parameter
\begin{equation}
\eta\equiv\frac{r_h^4}{\omega ^4}.
\end{equation}
The leading order differential equation is a Bessel differential equation. By using $\omega=i\,\zeta_n$, the regularity condition at $r=r_h$, that is for large $y$, depends on the sign of $n$.
For $n>0$, the selected leading order solution is $H^{(1)}_{\Delta -2}(y)$, whereas for $n<0$ the selected one is $H^{(2)}_{\Delta -2}(y)$. We present the results for $n>0$, the discussion for $n<0$ is analogous, up to the replacement of Hankel $H^{(1)}$ functions with Hankel $H^{(2)}$ functions. Hence, we write
\begin{equation}
\chi^{>}(y)=H_{\Delta -2}^{(1)}(y)+\sum_{j\ge 1}\eta^j\,\chi_j^{>}(y).
\end{equation}
The first subleading correction reads
\begin{align}
\chi_1^{>}(y)&=\biggl[\left(\frac{1}{5} (\Delta -4) y^4+\frac{1}{10} (\Delta -4) (\Delta -3) \Delta  y^2\right) H_{\Delta -2}^{(1)}(y)\\
&\quad-\left(\frac{y^5}{5}+\frac{1}{10} (\Delta -4) \Delta  y^3+\frac{1}{5} \Delta  \left(\Delta ^3-8 \Delta ^2+19 \Delta -12\right) y\right) H_{\Delta -1}^{(1)}(y)\biggr]\nonumber.
\end{align}
In general, $\chi_j^{>}(y)$ is of the form
\begin{equation}
\chi_j^{>}(y)=p_{j,1}(y)\,H_{\Delta-1}^{(1)}(y)+p_{j,2}(y^2)\,H_{\Delta-2}^{(1)}(y),
\end{equation}
for some polynomials $p_{j,1}(y),p_{j,2}(y)$, where $p_{j,1}(y)$ has degree $\left(10 j-1-(-1)^j\right)/2$ for $j\ge 1$, and $p_{j,2}(y)$ has degree $\left(10 j-1+(-1)^j\right)/2$ for $j\ge 0$.

To solve the inhomogeneous differential equation, the following identities are useful:
\begin{align}
&\frac{\mathrm{d}}{\mathrm{d}y}\left(y^{2i-1}\,H^{(1)}_{\Delta-1}(y)\right)=y^{2 i-1} H_{\Delta -2}^{(1)}(y)-(\Delta -2 i) y^{2 i-2} H_{\Delta -1}^{(1)}(y),\\
&\frac{\mathrm{d}}{\mathrm{d}y}\left(y^{2i}\,H^{(1)}_{\Delta-2}(y)\right)=(\Delta +2 i-2) y^{2 i-1} H_{\Delta -2}^{(1)}(y)-y^{2 i} H_{\Delta -1}^{(1)}(y),\\
&\frac{\mathrm{d}^2}{\mathrm{d}y^2}\left(y^{2i-1}\,H^{(1)}_{\Delta-1}(y)\right)=(4 i-3) y^{2 i-2} H_{\Delta -2}^{(1)}(y)+\\
&\qquad\qquad\qquad\qquad\qquad\left[(2 i-\Delta ) (-\Delta +2 i-1) y^{2 i-3}-y^{2 i-1}\right] H_{\Delta -1}^{(1)}(y),\nonumber\\
&\frac{\mathrm{d}^2}{\mathrm{d}y^2}\left(y^{2i}\,H^{(1)}_{\Delta-2}(y)\right)=(1-4 i) y^{2 i-1} H_{\Delta -1}^{(1)}(y)+\\
&\qquad\qquad\qquad\qquad\quad\left[(\Delta +2 i-2) (\Delta +2 i-3) y^{2 i-2}-y^{2 i}\right] H_{\Delta -2}^{(1)}(y),\nonumber
\end{align}
for every $i\ge 1$.

Finally, to find the Fourier coefficients in large $n$ expansion for the Euclidean correlator, we expand the Hankel functions around $y=0$, and we take the ratio of the coefficients in front of the behaviours $y^{\pm(\Delta-2)}$.

The result is an asymptotic approximation of $\widetilde{G}_E(\zeta_n, 0)$ in $1/n$ which we denote by $\widetilde{G}^\text{asy}_E(\zeta_n, 0)$. The first few results are as follows
\bea
\widetilde{G}_E(\zeta_n, 0) \sim  \widetilde{G}^\text{asy}_E(\zeta_n, 0) &=& \beta^{4-2\Delta}\sum_{m=0}^\infty c_m n^{2\Delta-4-4m},\label{asymptoticansatz}\\
c_0 &=&-\frac{2\pi \csc(\pi\Delta)}{\Gamma(\Delta-2)^2} \pi^{2\Delta -4},\\
c_1 &=& -\frac{\pi \csc(\pi \Delta)\Gamma(1+\Delta)}{20\Gamma(\Delta - 4)\Gamma(\Delta-2)^2}\pi^{2\Delta - 4}.
\eea
We have performed the computation up to order $\eta^{69}$, the result of which can be found in the notebook attached to the submission. In the following subsection we perform a resurgence analysis of this asymptotic series. The coefficients grow as $(4m)!$, however, we find that there are no non-perturbative contributions missing from \eqref{asymptoticansatz} in the sense that applying Borel resummation to \eqref{asymptoticansatz} gives the exact result, $\widetilde{G}_E(\zeta_n, 0) = \mathcal{S}\widetilde{G}^\text{asy}_E(\zeta_n, 0)$.

\subsection{Resurgence analysis of the $1/n$ expansion}\label{sec:resurgence}
In this subsection, we turn to the asymptotic expansion in $1/n$ of $\widetilde{G}_E(\zeta_n,0)$, and we show, via a resurgent analysis, that all the non-perturbative corrections of the transseries representing  $\widetilde{G}_E(\zeta_n,0)$ are turned off. 

We start with \eqref{asymptoticansatz}, and reorganise as follows
\be
\widetilde{G}^\text{asy}_E(\zeta_n, 0) =  \beta^{4-2\Delta}n^{2\Delta - 4}\sum_{p=0}^\infty \frac{\mathfrak{a}_p}{n^p},
\ee
where $\mathfrak{a}_{4m} = c_m$ and $\mathfrak{a}_{4m+1} = \mathfrak{a}_{4m+2} = \mathfrak{a}_{4m+3} = 0$, and where the $c_m$ coefficients are analytically computable by the method detailed in section \ref{sec:1onnmethod}. Then, we make a `factorial-over-power' ansatz at large $p$,
\be
\mathfrak{a}_p = \sum_{j=1}^4\frac{\Gamma(p+\beta_0)}{\left(w_j\right)^{p+\beta_0}} \mathfrak{b}^{(j)}, \qquad w_j = A e^{\frac{i\pi}{4}(1+2j)}, \qquad j=1,2,3,4,\label{apansatz}
\ee
where $\beta_0, A$ are parameters to be determined, and $\mathfrak{b}^{(j)}$ are series in $1/p$.\footnote{We thank In\^es Aniceto for suggesting this pattern.}
Fixing $\mathfrak{a}_{4m+1} = \mathfrak{a}_{4m+2} = \mathfrak{a}_{4m+3} = 0$ we get the relations
\be\label{relationsb0}
\mathfrak{b}^{(2)}_0 = e^{-\frac{3}{2}\pi i \beta_0}\mathfrak{b}^{(1)}_0,\quad \mathfrak{b}^{(3)}_0 = e^{-\pi i \beta_0}\mathfrak{b}^{(1)}_0, \quad \mathfrak{b}^{(4)}_0 = e^{-\frac{1}{2}\pi i \beta_0}\mathfrak{b}^{(1)}_0.
\ee
Reinserting these relations inside the ansatz \eqref{apansatz}, we find, at leading order,
\be
c_m = \mathfrak{a}_{4m} =  (-1)^m\frac{\Gamma(4m+\beta_0)}{A^{4m+\beta_0}} e^{-\frac{3}{4} \pi i \beta_0}  4 \mathfrak{b}^{(1)}_0 + \ldots,
\ee
which matches observed behaviour of the $c_m$ coefficients. Then, a useful ratio is
\be
\frac{c_{m+1}}{c_m m^4} = -\frac{256}{A^4} - \frac{128(3+2 \beta_0)}{A^4 m} + \ldots
\ee
which allows us to determine $A,\beta_0$ by comparing to computed $c_m$ values. Using Richardson extrapolation for the first 100 $c_m$ coefficients at $\Delta = 11/4$, we find 
\be
A = \sqrt{2}\pi, \qquad \beta_0 = 0,
\ee
which determines the four leading terms in \eqref{apansatz}.

Now, we can consider each term from \eqref{apansatz} at leading order, in the Borel plane. We will consider the sum from $p=1$ only, which won't affect our consideration of the singularities. We consider the Borel transform of each of the series
\begin{equation}
\sum_{p=1}^{\infty}\frac{\Gamma(p+\beta_0)}{(w_j)^{p+\beta_0}}\mathfrak{b}_0^{(j)}\frac{1}{n^p}
\end{equation}
for $j=1,2,3,4$,
\be
\mathcal{B}_j(\xi) = \sum_{p=1}^\infty \frac{\Gamma(p+\beta_0)}{\left(w_j\right)^{p+\beta_0} } \mathfrak{b}^{(j)}_0 \frac{\xi^p}{p!} =  \mathfrak{b}^{(j)}_0\sum_{p=1}^\infty \frac{1}{\left(w_j\right)^{p}} \frac{1}{p} \xi^p = -\mathfrak{b}_0^{(j)}\log\left(1-\frac{\xi}{w_j}\right),
\ee
and their Laplace transforms in a direction $\theta$:
\be
\mathcal{S}_j^\theta = \int_0^{e^{i\theta}\infty}e^{-\xi} \mathcal{B}_j\left(\frac{\xi}{n}\right) d\xi = \int_0^{\infty}e^{-e^{i\theta}\xi} \mathcal{B}_j\left(\frac{e^{i\theta}\xi}{n}\right) e^{i\theta}d\xi.
\ee
The differences of the lateral summations on either side of the log branch points will pick up the discontinuities and generate non-perturbative terms. Let $\theta_j = \arg(w_j)$. Then,
\begin{equation}
\begin{aligned}
\mathcal{S}_j^{\theta_j+\epsilon} - \mathcal{S}_j^{\theta_j-\epsilon} &= -\mathfrak{b}_0^{(j)}e^{i\theta_j}\int_0^{\infty}e^{-e^{i\theta_j}\xi} \left(\log\left(1-\frac{e^{i\theta_j+i\epsilon}\xi}{n w_j}\right)-\log\left(1-\frac{e^{i\theta_j-i\epsilon}\xi}{n w_j}\right)\right)d\xi\\
&= -\mathfrak{b}_0^{(j)}e^{i\theta_j}\int_0^{\infty}e^{-e^{i\theta_j}\xi} \left(\log\left(1-\frac{e^{i\epsilon}\xi}{n A}\right)-\log\left(1-\frac{e^{-i\epsilon}\xi}{n A}\right)\right)d\xi\\
&= 2\pi i\mathfrak{b}_0^{(j)}e^{i\theta_j}\int_{nA}^{\infty}e^{-e^{i\theta_j}\xi} d\xi\\
&= 2\pi i\mathfrak{b}_0^{(j)} e^{-w_j n}.
\end{aligned}
\end{equation}
Thus we conclude that the appropriate transseries ansatz that completes the perturbative sector for $\widetilde{G}_E(\zeta_n, 0)$ is 
\be
\widetilde{G}_E(\zeta_n, 0) =  \beta^{4-2\Delta}n^{2\Delta - 4}\left(\sum_{p=0}^\infty \frac{\mathfrak{a}_p}{n^p} + \sum_{j = 1}^4 \sigma_j e^{-w_j n} P_j + (\text{subleading})\right),
\ee
where the $P_j$ are perturbative series in $1/n$ and the final term encodes subleading non-perturbative sectors.
By inserting this expansion in \eqref{fourierseries}, one would find that for each non-perturbative contribution of the form $e^{-w n}$ with non-zero transseries parameter $\sigma$, there is a pair singularities in the complex $z$ plane at $z=e^{\pm w}$, where the one with $|z|>1$ is associated with the sum over the positive modes \eqref{fkz}, and the one with $|z|<1$ is associated with the sum over the negative modes $f_0(1/z)$. These are precisely the bouncing singularity locations, \eqref{bouncingtau}, since $\frac{2\pi}{\beta}\tau_j = w_j$.

\begin{figure}[h!]
\centering
\includegraphics[width=0.9\columnwidth]{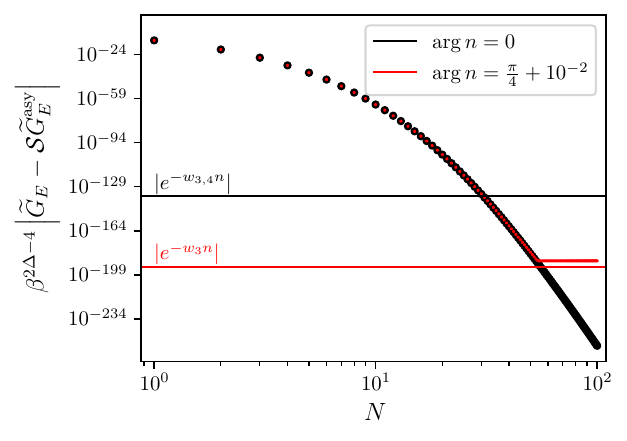}
\caption{Absence of non-perturbative contributions in the $1/n$ asymptotic expansion for $\widetilde{G}_E(\zeta_n, 0)$ \eqref{asymptoticansatz} when $n = 100$. The black points show the residual of the perturbative sector truncated to order $N$, while the black solid line shows the expected scale of non-perturbative terms, which are clearly absent. This shows all transseries parameters $\sigma_j$ vanish and hence the exact result is given by the Borel resummation of the perturbative sector only. Additionally, we show the case $n=100 e^{i\left(\frac{\pi}{4}+10^{-2}\right)}$ in red, corresponding to the result after crossing the Stokes line at $\arg{n} = \pi/4$, where the non-perturbative sector corresponding to bouncing singularities is turned on.}
\label{fig:resurgence_residual}
\end{figure}

However, for $n>0$ we have $\sigma_j = 0$. Thus the asymptotic expression $G_E^\text{asy}$ consists of the perturbative $1/n$ series only. Thus the non-perturbative contributions in the transseries associated to the bouncing singularities are turned off in this Stokes sector. Analytically continuing $n$ to complex values can lead to these terms being turned on. The evidence for this is numerical -- in figure \ref{fig:resurgence_residual} we show the difference between the exact numerical result for $\widetilde{G}_E(\zeta_n, 0)$ and the Borel-Pad\'e resummation of the $1/n$ series truncated to order $N$. For real $n$ the residual is far smaller than the scale of non-perturbative corrections, showing that $\sigma_j = 0$ for all $j$ in this sector. After crossing the Stokes line at $\arg{n} = \pi/2$ the non-perturbative corrections are present.

\section{Analytic structure from the Fourier series} \label{sec:analyticity}
The asymptotic growth of the Fourier coefficients \eqref{introasygrowth} at large $n$ is due to the singularity at the coincident point, i.e. at $\tau = 0$ (plus images), where the correlator has the singular behaviour, 
\be
G_E(\tau, k) \sim \frac{2(\Delta-2)\Gamma\left(\Delta - \frac{3}{2}\right)}{\sqrt{\pi} \Gamma(\Delta-2)}\frac{1}{\tau^{2\Delta-3}} \qquad \tau \to 0^+, \label{CoincindentPointSingularity}
\ee
consistent with the behaviour expected from CFT. This singularity sits on the edge of the regions of convergence for the $n>0$ and $n<0$ Fourier sums, preventing the convergence of the combined sum generally for $\tau \in \mathbb{C}$. In this section we use two complementary approaches to analytically continuing $G_E(\tau, k=0)$ to $\tau \in \mathbb{C}$ using the Fourier series data.

In what follows it will be helpful to define the coordinate
\be
z = e^{\frac{2\pi i}{\beta}\tau},
\ee
and split the Fourier series into the constant, upper-half plane and lower-half plane sums as follows,
\be\label{Ginf}
G_E(\tau, k) = \frac{1}{\beta}\left(\widetilde{G}(0,k) + f_k(z) + f_k(1/z)\right),
\ee
where we have defined
\begin{equation}\label{fkz}
f_k(z)=\sum_{n\ge 1}\widetilde{G}_E(\zeta_{n>0},k)z^n.
\end{equation}
Thus our attention will be on computing $f_k(z)$ in the complex $z$ plane, and analytically continuing the Taylor series \eqref{fkz} beyond the coincident point singularity at $z=1$. In this way \eqref{Ginf} can be evaluated for $z\in \mathbb{C}$ and thus $\tau \in \mathbb{C}$.

\subsection{Pad\'e approximants}\label{sec:pade}
\begin{figure}[h!]
\centering
\includegraphics[width=0.9\columnwidth]{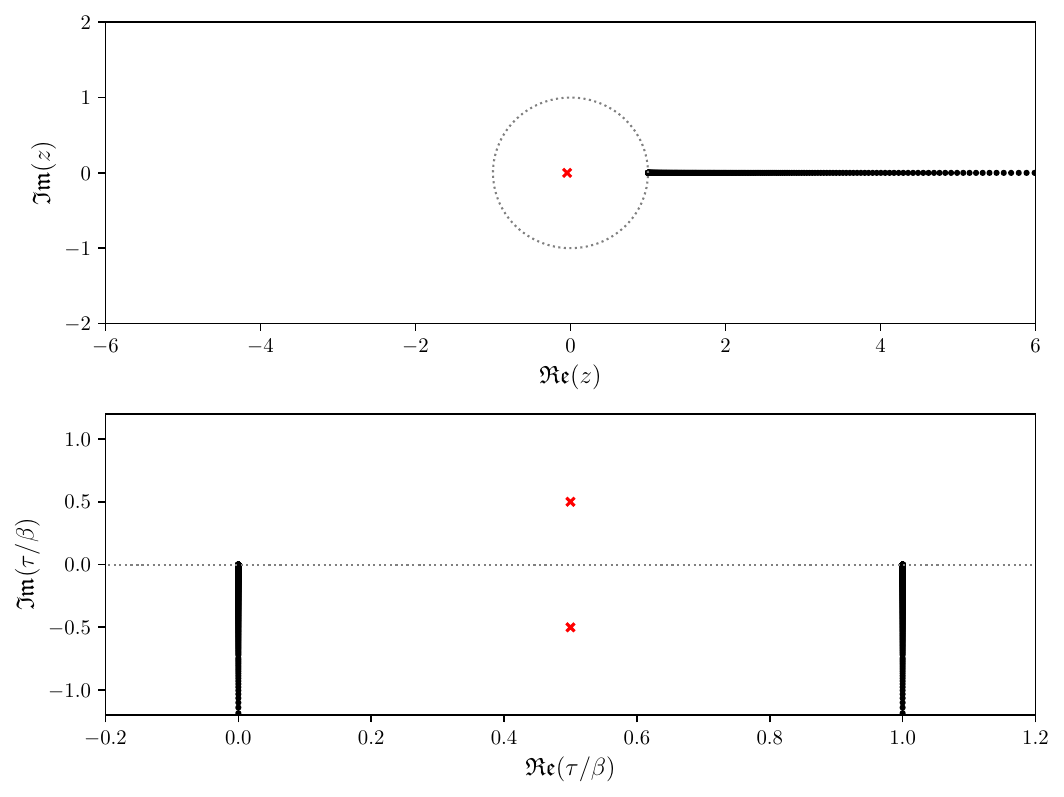}
\caption{Analytic structure of $f_0(z)$ explored using Pad\'e approximants. Black points show poles of the diagonal Padé approximant for 800 terms in the Fourier series of the holographic calculation at $\Delta = 11/4$, in the $z$-plane (upper panel) and the $\tau$-plane (lower panel). The results are consistent with analyticity of $f_0(z)$ for $z\in \mathbb{C}\setminus [1,\infty[$, equivalent to analyticity of $G_E(\tau, k=0)$ in the strip $0 < \re(\tau) < \beta$. The red crosses correspond to `bouncing singularity' locations $z_\pm = -e^{\pm \pi}$, i.e. \eqref{bouncingtau}, where the function appears regular. We note that the findings are similar when $k\neq 0$.}
\label{fig:zplane}
\end{figure}
In our first approach we will compute connection coefficients by numerically computing the Wronskians appearing in \eqref{connformulas}. In particular, the Heun functions and their derivatives appearing in these expressions can be evaluated to arbitrary precision. This gives us the Fourier coefficients numerically. To extend the radius of convergence of $f_k(z)$ beyond the disk $|z|<1$ we utilise diagonal Pad\'e approximants. An accumulation of poles in the Pad\'e approximant indicate the singular locations of the underlying function.

In figure \ref{fig:zplane} we show the locations of poles of the Pad\'e approximant in the $z$ and $\tau$ planes for $\Delta = 11/2$, $k=0$, using 800 terms. The results are consistent with the conicident point singularity \eqref{CoincindentPointSingularity} at $z=1$ and analyticity in the strip $0 < \re(\tau) < \beta$. In particular, we have marked the locations of the bouncing singularities \eqref{bouncingtau}, and there is no indication of any singularity present there from the analytic continuation.

Later in section \ref{sec:OPE} we show that the Pad\'e approximants are a good way of computing subleading terms in the OPE expansion of the function around $\tau = 0$ (and fully consistent with the approach using an asymptotic expansion of Fourier coefficients).

\subsection{Instanton approximation}\label{sec:instanton}
In this subsection we use the instanton approximation method to compute the connection coefficients through \eqref{Fouriercoeffgauge}. The advantage is that we can obtain analytic expressions for the Fourier series coefficients in the limiting case of integer $\Delta$, where it becomes straightforward to analytically continue $f_k(z)$ to $z\in\mathbb{C}\setminus\{1\}$.\footnote{We remark that the case of integer $\Delta$ was also studied in an analogous way in \cite{Jia:2025jbi}.}

Firstly, as detailed in Appendix \ref{app:gauge}, the quantities $a$ and $F(X)$ are expressed as series in $X$, but the overall structure of $\widetilde{G}_E(\zeta_n,k)$ is not affected by the instanton truncation order. In particular, it is not restrictive to consider the leading instanton order of \eqref{Fouriercoeffgauge} to study the singularity structure of the correlator. More precisely, the singularity structure is associated with the large $n$ expansion of $\widetilde{G}_E(\zeta_n,k)$. At each instanton order, the structure of the large $n$ expansion of $a^2$ and $\partial_{a_X}F(X)$ remains the same. Indeed, $a^2$ admits an expansion of the form $a_{0,n}\,n^2+a_{1,n}+\mathcal{O}(1/n^2)$, and $\partial_{a_X}F(X)$ admits an expansion of the form $b_{0,n}+\mathcal{O}(1/n^2)$, for some coefficients $a_{0,n}, a_{1,n}, b_{0,n}$ which change adding more instanton corrections.

Secondly, the Heun connection problem \eqref{connectionn>0} with the connection coefficients specified in Appendix \ref{app:gauge} is analogous to the connection problem of a hypergeometric equation, with one of the singularities having index equal to $2a$ \cite{Bonelli:2022ten, Arnaudo:2024rhv}. At leading instanton order, the analogy works perfectly, since the factor $e^{\partial_{a_X}F(X)}$ is absent, whereas the factor $X^{2-\Delta}$ appears because the connection problem involves the singularity at $v=X$ instead of the one at $v=1$.

When working with integer values of $\Delta$, the standard connection formulae do not apply straightforwardly.\footnote{In particular, the basis of local solutions of the differential equation has to be supplemented by logarithmic solutions around the point having an integer index.} The logarithmic connection formulae for the hypergeometric equation are known, and can be found for example in \cite{wangguo}. In our case, using Equation (9) on page 168 in \cite{wangguo}, at leading instanton order, the integer $\Delta$ version of \eqref{Fouriercoeffgauge} is\footnote{We again restrict to $k=0$ for simplicity, but the reasoning holds for any value of $k$.}
\begin{equation}
\widetilde{G}_E(\zeta_{n>0},k=0)=\frac{2\Delta-4}{r_h^{4-2\Delta}}\frac{X^{2-\Delta}}{\Gamma (\Delta -2)}\frac{\prod_{\sigma=\pm}\Gamma \left(\frac{\Delta-1+n}{2} +\sigma\sqrt{\frac{(\Delta -2)^2-1}{4}-\frac{3 n^2}{4}}\right)}{\prod_{\sigma=\pm}\Gamma \left(\frac{3+n-\Delta}{2}+\sigma\sqrt{\frac{(\Delta -2)^2-1}{4}-\frac{3 n^2}{4}}\right)}.
\end{equation}
For integer values of $\Delta$, the ratio of $\Gamma$ functions simplify to a polynomial in $n$ of degree $2\Delta-4$, $p_{2\Delta-4}(n)$. Finally, when computing $f_0(z)$ as in \eqref{fkz}, the result is a rational function in $z$ of the form $p_{2\Delta-3}(z)/(1-z)^{2\Delta-3}$.\footnote{The degree of the polynomial in the numerator can decrease if the constant term in $p_{2\Delta-4}(n)$ is not there.}

For example, at $\Delta=4$ and at leading instanton order, we find 
\begin{equation}
\begin{aligned}
f_0(z) &= \sum_{n\ge 1}\widetilde{G}_E(\zeta_{n>0},k=0)z^n|_{\Delta=4}\\
&=16r_h^4\sum_{n\ge 1} \left(n^4-\frac{5 n^2}{4}+\frac{1}{4}\right)  z^n=\frac{4r_h^4\, z^2 \left(z^3-5 z^2+55 z+45\right)}{(1-z)^5}.
\end{aligned}
\end{equation}
We remark that the infinite sum converges for $|z|<1$. However, in this case, the analytic continuation for $z\in\mathbb{C}\setminus\{1\}$ is straightforward since the result is a rational function.

Using formula \eqref{Ginf} blindly, we would find (at leading order in the instanton expansion) $G_E(\tau,k=0)=0$, which is a consequence of $p_{2\Delta-4}(n)$ being in fact a (quadratic) polynomial in $n^2$. However, the crucial point is that the rigorous result does not coincide with the constant function equal to 0, but has a distributional nature, with a divergence at $z=1$. Indeed, let us reintroduce the variable $\tau$, and write $G_E(\tau,0)$ (at leading order in the instanton expansion) as
\begin{equation}
G_E^{\text{leading\ inst.}}(\tau,k=0)=\frac{16r_h^4}{\beta}\sum_{n\in\mathbb{Z}}\left(n^4-\frac{5n^2}{4}+\frac{1}{4}\right)e^{\frac{2\pi i n}{\beta}\tau}.
\end{equation}
Then, we can use
\begin{equation}
\sum_{n\in\mathbb{Z}}e^{\frac{2\pi i n}{\beta}\tau}=\beta\sum_{l\in\mathbb{Z}}\delta(\tau-l\,\beta),
\end{equation}
and, for any $j\in \mathbb{Z}_{\ge 1}$,
\begin{equation}
\partial_{\tau}^j\,e^{\frac{2\pi i n}{\beta}\tau}=\left(\frac{2\pi i n}{\beta}\right)^je^{\frac{2\pi i n}{\beta}\tau},
\end{equation}
from which
\begin{equation}
n^j\,e^{\frac{2\pi i n}{\beta}\tau}=\left(\frac{\beta}{2\pi i}\right)^j\partial_{\tau}^j\,e^{\frac{2\pi i n}{\beta}\tau}.
\end{equation}
Applied to our polynomial, we find
\begin{equation}
\begin{aligned}
G_E^{\text{leading\ inst.}}(\tau,k=0)&=\frac{16r_h^4}{\beta}\left[\left(\frac{\beta}{2\pi}\right)^4\partial_{\tau}^4+\frac{5}{4}\left(\frac{\beta}{2\pi}\right)^2\partial_{\tau}^2+\frac{1}{4}\right]\sum_{n\in\mathbb{Z}}e^{\frac{2\pi i n}{\beta}\tau}\\
&=16r_h^4\left[\left(\frac{\beta}{2\pi}\right)^4\partial_{\tau}^4+\frac{5}{4}\left(\frac{\beta}{2\pi}\right)^2\partial_{\tau}^2+\frac{1}{4}\right]\sum_{l\in\mathbb{Z}}\delta(\tau-l\,\beta)\\
&=\sum_{l\in\mathbb{Z}}\left[\frac{\beta^4\,r_h^4}{\pi^4}\delta^{(4)}(\tau-l\,\beta)+\frac{5\beta^2\,r_h^4}{\pi^2}\delta^{(2)}(\tau-l\,\beta)+4\,r_h^4\delta(\tau-l\,\beta)\right],
\end{aligned}
\end{equation}
which, in the $z$ variable, is only singular at $z=1$. The outcome, at leading instanton order, is therefore a set of contact terms. Whilst these are scheme dependent, it once more highlights the distributional nature of the correlation function, and the lack of further singularities in the complex $\tau$ plane.

As discussed, this is not the full result for $G_E(\tau,k=0)$, since this computation takes into account the leading instanton order only, but the singularity structure in the $z$ complex plane is not affected by adding more orders. Adding instanton contributions changes the polynomial $p_{2\Delta-4}(n)$ into a rational function in $n$, for which the Fourier series cannot be evaluated in closed form, however the radius of convergence of the Fourier series will remain the same, i.e. corresponding to the singularity at $z=1$.

\section{Extracting OPE coefficients} \label{sec:OPE}

In this section we extract the stress-tensor and double-trace coefficients from the small $\tau$ expansion of the Fourier series at $k=0$. As the results are automatically periodic, no further steps are required to obtain the double trace contributions.

In the first instance, it is convenient to work directly with the asymptotic expansion of the Fourier series coefficients in $1/n$, \eqref{asymptoticansatz}. Recall that this expansion is divergent, but does not require any additional non-perturbative contributions, as shown in section \ref{sec:resurgence}. Thus to make use of it only requires performing a Borel resummation. Treating the divergent series as formal ones, we swap the $m$ sum in \eqref{asymptoticansatz} with the $n$ sum \eqref{fkz}, getting
\be
f_0^\text{asy}(z) = \beta^{4-2\Delta}\sum_{m=0}^\infty c_m\text{Li}_{4+4m-2\Delta}(z).\label{polylogs}
\ee
Each term in this sum can be continued to the entire complex plane, where it is analytic except for a singularity at $z=1$. We have for $\tau >0$,
\begin{equation}
\begin{aligned}
\text{Li}_{4+4m-2\Delta}(z) + \text{Li}_{4+4m-2\Delta}(1/z) &= -2\sin(\pi\Delta)\Gamma(2\Delta - 3-4m)\left(\frac{2\pi\tau}{\beta}\right)^{-2\Delta +3+ 4m}\\
&+\sum_{q=0}^\infty \frac{2(-1)^q}{(2q)!}\zeta(-2\Delta + 4m+4-2q)\left(\frac{2\pi\tau}{\beta}\right)^{2q}.
\end{aligned}
\end{equation}
This should be modified if $2\Delta$ is an integer, because there will also be logarithmic contributions.
Then, from \eqref{Ginf}, we have
\bea
G_E^\text{asy}(\tau, 0) &=& \beta^{3-2\Delta}\sum_{m=0}^\infty \frac{\pi\, c_m }{\cos(\pi\Delta)\Gamma(4m+4-2\Delta)}\left(\frac{2\pi\tau}{\beta}\right)^{-2\Delta +3+ 4m}\label{OPEcm}\\
&&+ \frac{\widetilde{G}_E(0,0)}{\beta} +\beta^{3-2\Delta}\sum_{q=0}^\infty \frac{2(-1)^q}{(2q)!} \left[\sum_{m=0}^\infty c_m\zeta(-2\Delta + 4m+4-2q)\right]\left(\frac{2\pi\tau}{\beta}\right)^{2q},\nonumber
\eea
where in the last term we again exchanged the $m$ sum and the $q$ sum. The first line in \eqref{OPEcm} denotes the contribution of stress tensors, in which the $(4m)!$ divergence of the asymptotic expansion is removed. The second line is the double-trace sector, also determined by the same set of coefficients, but it remains divergent. Thus to evalute them we need to perform a Borel resummation. We denote the divergent sum,
\be
s_q = \sum_{m=0}^\infty c_m\zeta(-2\Delta + 4m+4-2q), \label{sqdef}
\ee
which in practice is resummed through Borel-Pad\'e for the first $N+1$ coefficients in the asymptotic expansion,
\be
\tilde{s}_q = \int_0^{\infty}e^{-y}\,\text{DiagonalPadé}\left[\sum_{m=0}^N\frac{c_m\,\zeta(4m-2\Delta+4-2q)}{(4m)!}y^{4m},\lfloor \frac{N}{2} \rfloor \right]\mathrm{d}y. \label{stilde}
\ee
As a reminder there are no non-perturbative corrections in this sector, and so this result is exact, as $N\to \infty$.

Thus our final expression for the dimensionless form of the OPE coefficients, as defined in \eqref{OPEintro}, are as follows
\bea
a_m &=& \frac{\pi\, c_m }{\cos(\pi\Delta)\Gamma(4m+4-2\Delta)}\left(2\pi\right)^{4m-2\Delta +3}, \label{acoeffs}\\
b_q &=& \frac{\widetilde{G}_E(0,0)}{\beta^{4-2\Delta}}\delta_{q0}+  \frac{2(-1)^q}{(2q)!} \tilde{s}_q\left(2\pi\right)^{2q},\label{bcoeffs}
\eea
where $c_m$ are the asymptotic expansion coefficients of the Fourier series \eqref{asymptoticansatz} which are analytically computable, and $\tilde{s}_q$ is determined by them using \eqref{stilde}. $\widetilde{G}_E(0,0)$ is given by \eqref{G00}.

\begin{figure}[h!]
\centering
\includegraphics[width=\columnwidth]{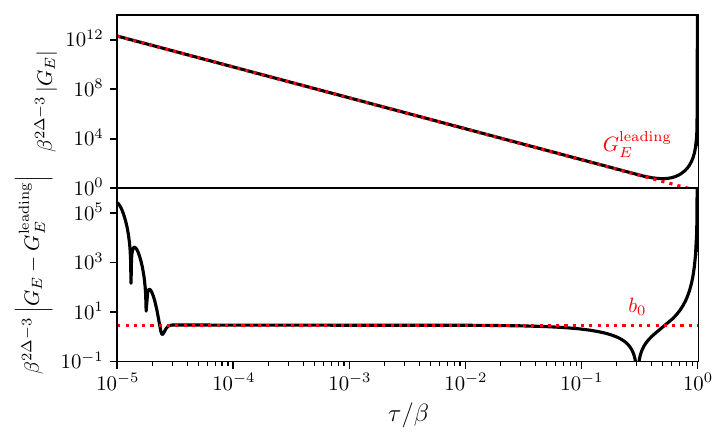}
\caption{The near-singularity behaviour of the Euclidean correlator at $k=0$, using OPE coefficients computed from asymptotic formulae (red dash) and  the Pad\'e approximant of 2000 terms in the Fourier series (black solid).  We work at $\Delta = 11/4$. \textbf{Upper panel:} Here we show the leading singular behaviour at $\tau  = 0$, and the behaviour \eqref{CoincindentPointSingularity}. \textbf{Lower panel:} After subtracting the leading singular behaviour the next OPE coefficient is $b_0$ corresponding to the leading double-trace sector term, which is constant in $\tau$. For small enough $\tau$ there is a residual corresponding to the breakdown of the fine cancellation of the leading singular behaviour, but this improves with more terms used in the Fourier series.}
\label{fig:subtraction}
\end{figure}

\begin{figure}[h!]
\centering
\includegraphics[width=0.9\columnwidth]{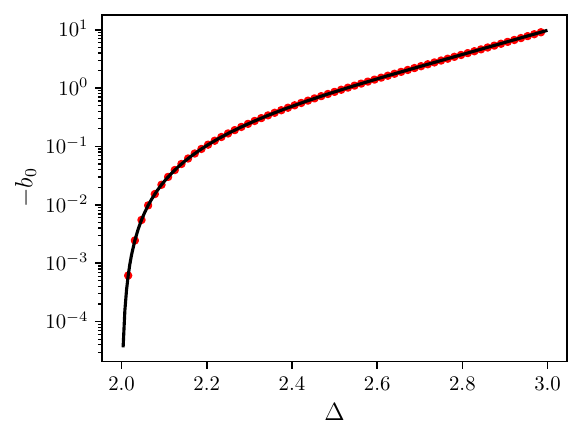}
\caption{The leading double-trace sector coefficient in the OPE expansion \eqref{OPEintro} at $k=0$, $b_0$, for a range of $\Delta$. The black curve is the coefficient computed from the Borel resummation of the asymptotic expansion of the Fourier series coefficients. The red dots are the same quantities but computed from the Pad\'e approximant of the Fourier series, subtracting the leading singularity, and fitting the remaining constant behaviour (a procedure shown in figure \ref{fig:subtraction} for the single case of $\Delta = 11/4$).}
\label{fig:b0delta}
\end{figure}

We can check the resulting value of $b_q$ against the Fourier series evaluation using Pad\'e approximants as discussed in section \ref{sec:pade}. The first 70 $c_m$ coefficients for generic $\Delta$ were computed in section \ref{sec:1onnmethod} and can be found in the attached notebook. We can then evaluate \eqref{bcoeffs}. For example, at $\Delta = 11/4$ we find
\be
b_0 \simeq -2.9593.
\ee
In figure \ref{fig:subtraction}, we show the correlator on the real $\tau$ line computing using Pad\'e approximants of the Fourier series. We also show the leading behaviour as $\tau\to 0$, $G_E^\text{leading}$, given by \eqref{CoincindentPointSingularity}. We then subtract $G_E^\text{leading}$, leaving the $\tau^0$ behaviour as the next OPE correction. As shown in the figure, the agreement with $b_0$ is excellent. We confirm the agreement between these two determinations of the OPE coefficient $b_0$ for varying $\Delta$ in $2<\Delta<3$ in figure \ref{fig:b0delta}.

\section{Outlook}\label{sec:outlook}
In this work we computed the Euclidean thermal two point function in holography, at fixed spatial momentum $\vec{k}$, based on its Fourier series representation \eqref{fourierseries}. The key advantages of this approach are that the Fourier series converges in the sense of distributions, it is manifestly periodic in $\tau$, and it can be efficiently computed with a variety of techniques including Pad\'e approximants, instanton expansions, and resurgence analysis. 

The manifestly distributional nature of the correlator in our treatment is consistent with QFT expectations. See also \cite{Kravchuk:2020scc, Kravchuk:2021kwe} for a related discussion in the CFT bootstrap context. Indeed, in section \ref{sec:instanton} we explicitly computed the contact terms which arise. Whilst these are scheme dependent, they are uniquely determined by the holographic computation given a choice of finite local counterterms, and so they are necessary to complete the result.

Because the result is manifestly periodic, its OPE expansion is complete; i.e. it not only contains stress-tensor contributions, but also the double-trace sector. The stress-tensor coefficients are given by the asymptotic expansion of the Fourier series coefficients, while the double traces are sums of them. The final expressions are given by \eqref{acoeffs} and \eqref{bcoeffs}. Note that the asymptotic expansion of the Fourier coefficients is divergent, and must be Borel resummed. This naturally explains the origin for a Borel resummation step in an approach to bootstrap double trace sectors from the stress tensors \cite{Buric:2025fye}. 

The Fourier series enabled us to study the analytic structure of the complex $\tau$ plane, here through analytic continuation using Pad\'e approximants and instanton expansions. Our results were consistent analyticity in the strip $0 < \re(\tau) < \beta$. In addition, we verified that the Borel resummation of the perturbative $1/n$ expansion of Fourier coefficients is exact. This is in contrast to the situation for the retarded Green's function \cite{Afkhami-Jeddi:2025wra, Giombi:2026kdz}, in which the difference between the exact result and the Borel resummed perturbative sector in $1/\omega$ for real $\omega$ encodes non-perturbative contributions corresponding to bouncing singularities. In the present Euclidean case, the non-perturbative sectors in the transseries also correspond to the bouncing singularities, but they are turned off. Analytically continuing from real $n$ to real $\omega$ crosses Stokes lines which activates them.\footnote{The Stokes lines are given by the rays $\arg(\omega) = \arg(\tau_j)$, where $\tau_j$ are the bouncing singularities \eqref{bouncingtau}, which are the same as angles for the asymptotic lines of QNMs and anti-QNMs \cite{Natario:2004jd, Cardoso:2004up}. See \cite{Dodelson:2025jff, Jia:2025jbi} for related results.}

We worked at fixed spatial momentum $\vec{k}$. Converting the result to fixed spatial separation $x$ requires the inverse transform of \eqref{spatialFT}. The integration will likely produce some divergences to deal with, which again find their roots in the distributional nature of the correlator. An analogous discussion holds for the global Schwarzschild-AdS case, for which one expects to encounter divergences when performing the infinite sum over the angular quantum numbers. Once this has been performed, the asymptotic expansion of the thermal circle Fourier series coefficients could be studied in the same way as in section \ref{sec:resurgence}, leading to the OPE formulae and non-perturbative sectors corresponding to bouncing singularities.  

Continuing to the Lorentzian case one expects to find light-cone and bulk-cone singularities, which have been studied by considering the eikonal or Regge limit \cite{Dodelson:2023nnr, Jia:2026pmv, Araya:2026shz}. An important development in this regard was established in \cite{Dodelson:2023vrw}, where the Wightman function was constructed in terms of a product over QNMs and anti-QNMs. The representation of the correlator as a sum over residues of QNMs and anti-QNMs is convergent, as also analysed in \cite{Arnaudo:2025uos}. The connection to our work can be understood through analytic continuation and considering a Sommerfeld-Watson transformation. For each strip $j\,\beta<\re(\tau)<(j+1)\,\beta$, $j\in\mathbb{Z}$, the Sommerfeld-Watson transformation is defined via a specific integration kernel; we describe the case $j=0$. By analytically continuing $n\in\mathbb{Z}$ in the complex plane, and reinterpreting it as the frequency of the perturbation $\omega\in\mathbb{C}$, we can rewrite \eqref{fourierseries} as
\begin{equation}
\begin{aligned}
&\frac{1}{\beta}\sum_{n=-\infty}^\infty   \widetilde{G}_E(\zeta_n, k)e^{i \zeta_n \tau} = \frac{1}{2\pi i}\int_\Gamma \frac{e^{\omega\tau}}{e^{\beta \omega}-1}\widetilde{G}_E(-i \omega,k) d\omega=\\
    &\frac{G_E(0,k)}{\beta} + \frac{1}{2\pi i}\int_{-\infty}^\infty \frac{e^{\omega\tau}}{e^{\beta \omega}-1}\left(\widetilde{G}_E(-i (\omega + i0),k)  - \widetilde{G}_E(-i (\omega-i0),k)\right) d\omega+ \text{arcs}\nonumber,
\end{aligned}
\end{equation}
where $\Gamma$ is a closed contour encircling the Matsubara frequencies and to get to the second expression we deformed the contour. The discontinuity integral contribution is finite in the strip $0 < \re(\tau) < \beta$ due to the exponential suppression of the prefactor. 
Thus the divergence of the Fourier sum, reflecting the distributional nature of the correlator, is reinterpreted as divergent contributions from the large arcs, whose divergent behaviour can be seen from the polynomial growth of $\widetilde{G}_E(-i\omega,k)$.

\section*{Acknowledgements}
It is a pleasure to thank Inês Aniceto, Alba Grassi, Cristoforo Iossa, Robin Karlsson and Elli Pomoni for useful discussions.
In addition we would like to thank Cristoforo Iossa, Robin Karlsson and IGAP Trieste for hospitality during the programme `Black Hole Perturbations and Holography' where this work was finalised.
PA is supported by the Royal Society grant URF\textbackslash R\textbackslash 231002, `Dynamics of holographic field theories'.
BW is supported by a Royal Society University Research Fellowship and in part by the STFC consolidated grant `New Frontiers In Particle Physics, Cosmology And Gravity'. 

\appendix

\section{Gauge theory details and notation}\label{app:gauge}

The connection formulae for the Heun equation were obtained in \cite{Bonelli:2022ten} using the language of $\mathcal{N}=2$ supersymmetric gauge theory.
In this Appendix, we briefly introduce the essential ingredients to write the connection coefficient expressions.

The most important object in this regard is the instanton partition function \cite{Nekrasov:2002qd, Nekrasov:2003rj}. By denoting with $\vec{Y}=(Y_1,Y_2)$ a pair of Young diagrams, with $\vec{a}=(a_1,a_2)$ the vacuum expectation value of the scalar in the vector multiplet, and with $\epsilon_1,\epsilon_2$ the parameters characterizing the $\Omega$-background, one can define the hypermultiplet and vector contributions
\begin{equation}
\begin{aligned}
z_{\text{hyp}} \left( \vec{a}, \vec{Y}, m \right)&=\prod_{k= 1,2} \prod_{(i,j) \in Y_k} \left[ a_k + m + \epsilon_1 \left( i - \frac{1}{2} \right) + \epsilon_2 \left( j - \frac{1}{2} \right) \right],\\
z_{\text{vec}} \left( \vec{a}, \vec{Y} \right)&=\prod_{i,j=1}^2\prod_{s\in Y_i}\frac{1}{a_i-a_j-\epsilon_1L_{Y_j}(s)+\epsilon_2(A_{Y_i}(s)+1)}\\
&\quad\times\prod_{t\in Y_j}\frac{1}{-a_j+a_i+\epsilon_1(L_{Y_i}(t)+1)-\epsilon_2A_{Y_j}(s)}.
\end{aligned}
\end{equation}
We always adopt the conventions $\epsilon_1=1$ and $\vec{a}=(a,-a)$.
By denoting with $m_1,m_2,m_3,m_4$ the masses of the four hypermultiplets, the monodromy parameters $a_0,a_X,a_1,a_{\infty}$ are defined as
\begin{equation}\label{gaugemasses}
\begin{aligned}
m_1&=-a_X-a_0,\quad
&m_2=-a_X+a_0,\\
m_3&=a_{\infty}+a_1,\quad
&m_4=-a_{\infty}+a_1.
\end{aligned}
\end{equation}
The position of the fourth singularity of the Heun equation, that we denoted with $X$, represents the instanton counting parameter $X=e^{2\pi i\tau}$, where $\tau$ is related to the gauge coupling by 
\begin{equation}
\tau=\frac{\theta}{2\pi}+i\frac{4\pi}{g_{\rm YM}^2}.
\end{equation}
The instanton part of the Nekrasov-Shatashvili (NS) free energy \cite{ns} is given as a power series in $X$ by
\begin{equation}
F(X)=\lim_{\epsilon_2\to 0}\epsilon_2\log\Biggl[(1-X)^{-2\epsilon_2^{-1}\left(\frac{1}{2}+a_1\right)\left(\frac{1}{2}+a_X\right)}\sum_{\vec{Y}}X^{|\vec{Y}|}z_{\text{vec}} \left( \vec{a}, \vec{Y} \right)\prod_{i=1}^4z_{\text{hyp}} \left( \vec{a}, \vec{Y}, m_i \right)\Biggr].
\end{equation}

The first few orders in the $X$ expansion of $F(X)$ read
\begin{equation}
F(X)=\frac{\left(4 a^2-4 a_0^2+4 a_X^2-1\right) \left(4 a^2+4 a_1^2-4 a_\infty^2-1\right)}{8-32 a^2}\,X+\mathcal{O}(X^2).
\end{equation}
The convergence properties of this series expansion are not rigorously studied in the NS limit, but in the case $X=1/2$ of our problem \eqref{Heundictio} all the expansions are expected to be convergent from the structure of the singularities in the $v$ plane\footnote{For results on the convergence of the instanton partition functions in more generic $\Omega$ backgrounds we refer to \cite{guillarmou2024conformal, Arnaudo:2022ivo}.}.

The gauge parameter $a$ parametrizes the composite monodromy around the points $z=0$ and $z=X$, and is expressed as a series expansion in the instanton counting parameter $X$, obtained by inverting the Matone relation \cite{Matone:1995rx}
\begin{equation}
u =-\frac{1}{4} - a^2 + a_X^2 + a_0^2 + X \partial_X F(X),
\end{equation}
where the parameter $u$ appearing in the differential equation \eqref{heunnormalform} is the accessory parameter.
The first few orders in the $x$ expansion of $a$ read
\begin{equation} 
a=\pm\left\{\sqrt{-\frac{1}{4}-u+a_X^2+a_0^2}+\frac{\bigl(\frac{1}{2}+u-a_X^2-a_0^2-a_1^2+a_{\infty}^2\Bigr)\Bigl(\frac{1}{2}+u-2a_X^2\Bigr)}{2(1+2u-2a_X^2-2a_0^2)\sqrt{-\frac{1}{4}-u+a_X^2+a_0^2}}X+\mathcal{O}(X^2)\right\}.
\end{equation}

With these notations, the connection coefficients can be written as \cite{Bonelli:2022ten}
\begin{equation}
\begin{aligned}
\mathcal{C}_{<,N.}&=X^{-a_0-a_X}e^{-\frac{1}{2}\partial_{a_0}F(X)-\frac{1}{2}\partial_{a_X}F(X)}\frac{\Gamma\left(1-2a_0\right)\Gamma\left(-2a_X\right)}{\Gamma\left(\frac{1}{2}-a_0-a_X+a\right)\Gamma\left(\frac{1}{2}-a_0-a_X-a\right)},\\
\mathcal{C}_{<,N.N.}&=X^{-a_0+a_X}e^{-\frac{1}{2}\partial_{a_0}F(X)+\frac{1}{2}\partial_{a_X}F(X)}\frac{\Gamma\left(1-2a_0\right)\Gamma\left(2a_X\right)}{\Gamma\left(\frac{1}{2}-a_0+a_X+a\right)\Gamma\left(\frac{1}{2}-a_0+a_X-a\right)},
\end{aligned}
\end{equation}
\begin{equation}
\begin{aligned}
\mathcal{C}_{>,N.}&=X^{a_0-a_X}e^{\frac{1}{2}\partial_{a_0}F(X)-\frac{1}{2}\partial_{a_X}F(X)}\frac{\Gamma\left(1+2a_0\right)\Gamma\left(-2a_X\right)}{\Gamma\left(\frac{1}{2}+a_0-a_X+a\right)\Gamma\left(\frac{1}{2}+a_0-a_X-a\right)},\\
\mathcal{C}_{>,N.N.}&=X^{a_0+a_X}e^{\frac{1}{2}\partial_{a_0}F(X)+\frac{1}{2}\partial_{a_X}F(X)}\frac{\Gamma\left(1+2a_0\right)\Gamma\left(2a_X\right)}{\Gamma\left(\frac{1}{2}+a_0+a_X+a\right)\Gamma\left(\frac{1}{2}+a_0+a_X-a\right)}.
\end{aligned}
\end{equation}

We also include the dictionary for the parameters entering the local solutions of the Heun equation in terms of the ones in \eqref{heunnormalform}:
\begin{equation}\label{Heun parameters}
\begin{aligned}
\alpha_1&=1-a_0-a_1-a_X+a_{\infty},\\
\alpha_2&=1-a_0-a_1-a_X-a_{\infty},\\
\gamma&=1-2a_0,\\
\delta&=1-2a_1,\\
\epsilon&=1-2a_X,\\
q&=\frac{1}{2}+X\left(a_0^2+a_1^2+a_X^2-a_{\infty}^2\right)-a_X-a_1\,X+a_0\left[2a_X-1+X\left(2a_1-1\right)\right]+\left(1-X\right)u.
\end{aligned}
\end{equation}

\bibliographystyle{ytphys}
\bibliography{refs}

\end{document}